\documentclass[12pt]{article}
\usepackage{epsfig}

\newcommand{\ls}{{\stackrel{\textstyle <}{_\sim}}}
\newcommand{\msun}{{M_{\odot}}}

\newcommand{\eos}{equation of state~}
\newcommand{\eoss}{equations of state~}

\newcommand{\eosp}{equation of state}

\newcommand{\fmmt}{{\rm fm}^{-3}}
\newcommand{\mevt}{{\rm MeV/fm}^3}

\newcommand{\mev}{{\rm MeV}}

\newcommand{\bag}{{B^{1/4}}}
\newcommand{\gcmt}{{\rm g/cm}^3}

\newcommand{\secm}{{\rm s}^{-1}}
\newcommand{\kFn}{{k_{F_n}}}
\newcommand{\fmmo}{{\rm fm}^{-1}}
\newcommand{\ergs}{{\rm erg/s}}
\newcommand{\AaA}{{\it Astron. {\&} Astrophys.~}}
\newcommand{\ApJ}{{\it Astrophys. Journal~}}
\newcommand{\NP}{{\it Nucl. Phys.} }
\newcommand{\PL}{{\it Phys. Lett.} }
\newcommand{\PR}{{\it Phys. Rev.} }
\newcommand{\PRL}{{\it Phys. Rev. Lett.} }
\newcommand{\etal}{{\it et al\/}\ }
\newcommand{\jpg}{{\it J. Phys. G: Nucl. Part. Phys.} } 

\begin{document}

%............... my section and subsection commands.........
\newcounter{sctn}
\newcounter{subsctn}[sctn]
\newcommand{\sctn}[1]{~\\ \refstepcounter{sctn} {\bf \thesctn~~ #1} \\ }
\newcommand{\subsctn}[1]
{~\\ \refstepcounter{subsctn} {\bf \thesctn.\thesubsctn~~ #1}\\}

%.................heading definitions.................
\newcommand{\dateofdoc}{\today}

\newcommand{\und}{\begin{flushright} FW-UNDAP-08-00 \\[5ex] \end{flushright}}

\newcommand{\tit}{\bf Strangeness in Neutron Stars}

\newcommand{\auth}{Fridolin Weber \\[6ex]}

\newcommand{\doe}{This work was supported by the Deutsche
  Forschungsgemeinschaft.}

\newcommand{\adr}{{University of Notre Dame \\ Department of Physics \\ 225
Nieuwland Science Hall \\ Notre Dame, IN 46556-5670, USA \\
http://nta0.lbl.gov/$\sim$fweber} \\[4ex]}

%%%%%%%%%%%%% FIRST TITLE PAGE%%%%%%%%%%%%%%%%%%%%
\begin{titlepage}
\und
\renewcommand{\thefootnote}{\fnsymbol{footnote}}
\setcounter{footnote}{0}
\vspace*{1.0cm}
\begin{center}
\begin{Large}
\tit \\[10ex]
\end{Large}
\renewcommand{\thefootnote}{\fnsymbol{footnote}}
\begin{large}
\auth
\end{large}
\adr
\dateofdoc \\[25ex]
\end{center}

%%%%%%% FOR NOTATION OF WHERE PRESENTED USE THE FOLLOWING
\begin{quote}
\begin{center} 
{Presented at the 5th International Conference \\ on \\ Strangeness in
Quark Matter (Strangeness 2000) \\ Berkeley, California, USA \\ July
20--25, 2000}
\end{center}
\end{quote}
\end{titlepage}

%............ Begin first page ....................
\renewcommand{\thefootnote}{\arabic{footnote}}
\setcounter{footnote}{0}
\begin{center}
\begin{Large}
\tit \\[4ex]
\end{Large}
\begin{large}
\auth
\end{large}
\adr
\end{center}
\vskip 1.0truecm

%............... ABSTRACT .................
\begin{abstract}
It is generally agreed on that the tremendous densities reached in the
centers of neutron stars provide a high-pressure environment in which
numerous novel particles processes are likely to compete with each
other. These processes range from the generation of hyperons to quark
deconfinement to the formation of kaon condensates and H-matter.
There are theoretical suggestions of even more exotic processes inside
neutron stars, such as the formation of absolutely stable strange
quark matter, a configuration of matter even more stable than the most
stable atomic nucleus, iron. In the latter event, neutron stars would
be largely composed of pure quark matter, eventually enveloped in a
thin nuclear crust. No matter which physical processes are actually
realized inside neutron stars, each one leads to fingerprints, some
more pronounced than others though, in the observable stellar
quantities. This feature combined with the unprecedented progress in
observational astronomy, which allows us to see vistas with remarkable
clarity that previously were only imagined, renders neutron stars to
nearly ideal probes for a wide range of physical studies, including
the role of strangeness in dense matter.
\end{abstract}

% \pacs{12.38.A, 12.38.M, 26.60, 82.60.F, 97.60.G, 97.60.J}

\newpage

\begin{Large}
\tit 
\end{Large}

\section{Introduction}\label{sec:intro}

\begin{figure}[tbh] 
\begin{center}
\leavevmode
\epsfig{figure=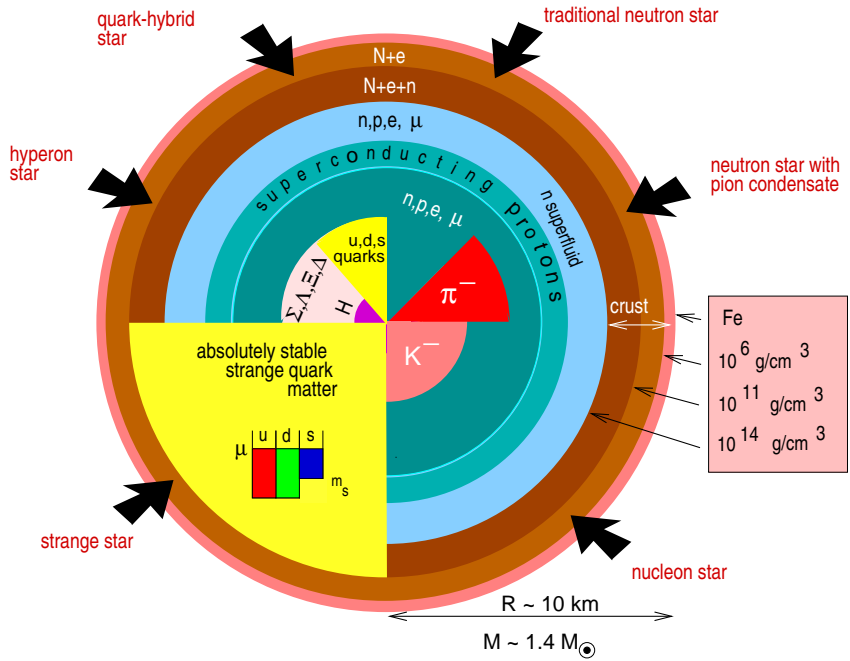,width=8.7cm,angle=-90}
\caption[]{Competing structures and novel phases of subatomic matter
predicted by theory to make their appearance in the cores ($R\ls
8$~km) of neutron stars. Corresponding models for the \eos are shown
in figure~\ref{fig:eos}.}
\label{fig:cross}
\end{center}
\end{figure}

\begin{figure}[tbh]
\begin{center}
\leavevmode 
\epsfig{figure=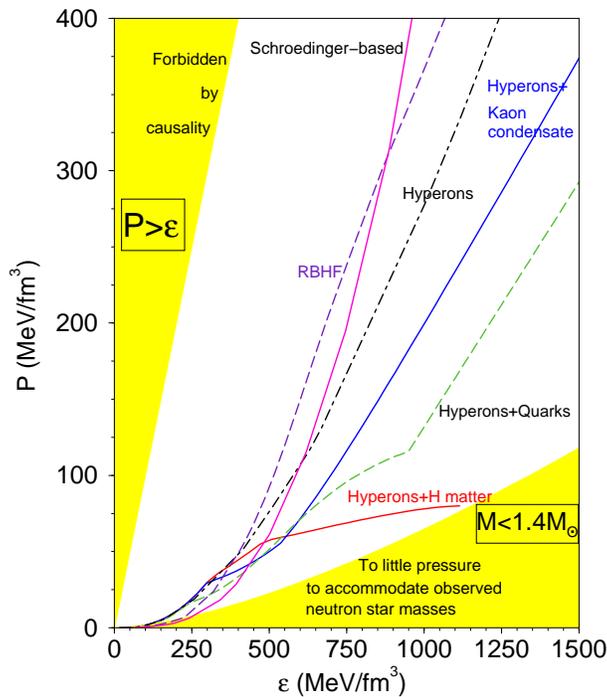,width=8.0cm,angle=0} 
\caption[Graphical illustration of equations of state]{Models for the equation 
  of state of ``neutron'' star matter
\protect{\cite{weber99:book,glen99:a,glen98:a}}.}
\label{fig:eos}
\end{center}
\end{figure}

Neutron stars are spotted as pulsars by radio telescopes and x-ray
satellites.  They are more massive (i.e. $\sim 1.5\, \msun$) than our
sun but are typically only about $\sim 10$ kilometers across so that
the matter in their centers is compressed to densities that are up to
an order of magnitude higher than the density of atomic nuclei
\cite{glen97:book,weber99:book}.  At such densities it is quite
plausible that numerous subatomic particle processes will compete with
each other and novel phases of matter -- like the quark-gluon plasma
being sought at the most powerful terrestrial particle colliders --
could exist in the center of neutron stars. Figure~\ref{fig:cross}
summarizes the possible scenarios graphically, with the associated
\eoss shown in figure~\ref{fig:eos}.
 The strangeness-carrying $s$ quark is likely to play a vital 
role inside neutron stars, for several of the building blocks of
matter may contain an $s$ quark as one of their constituents. Examples
are the hyperons $\Lambda, \Sigma, \Xi$, the $K^-$ meson, and the
H-dibaryon.  More than that, strangeness may also exist in the form of
unconfined $s$ quarks, which could populate, in chemical equilibrium
with $u$ and $d$ quarks and/or hadrons, extended regions inside
neutron stars.  

This paper aims at giving a brief overview of the possible
manifestations of strangeness inside neutron stars. It is most
noteworthy that these objects can nowadays be observed -- with
state-of-the-art radio telescopes (e.g.\ Arecibo, Jodrell Bank,
Parkes, VLA) and x-ray satellites (e.g.\ Chandra, HST, RXTE, XMM) --
with such a remarkable clarity that previously was only
imagined,\footnote{See, for instance, {\tt
http://www.r-clarke.org.uk/astrolinks\_radio.htm}, \hfill\break {\tt
http://pulsar.princeton.edu/}, {\tt http://heasarc.gsfc.nasa.gov/}.}
which enables us for the first time ever to seriously constrain the
properties of superdense matter from astrophysical observations.

\section{Hyperons}

Only in the most primitive conception, a neutron star is constituted
from neutrons.  At a more accurate representation, a such stars will
contain neutrons ($n$) and a small number of protons ($p$) whose
charge is balanced by leptons ($e^-$, $\mu^-$).  At the densities in
the interiors of neutron stars, the neutron chemical potential,
$\mu^n$, easily exceeds the mass (modified by interactions) of the
$\Lambda$ which then allows neutrons to transform into $\Lambda$'s.
Mathematically, this is expressed as $\mu^n = \mu^\Lambda$, which
indicates that for neutron Fermi momenta greater than $\kFn \sim 3 \,
\fmmo$, $\Lambda$ particles could make their appearance. Such Fermi
momenta correspond to densities of just $\sim 2 \rho_0$, with $\rho_0
=0.16\, \fmmt$ the baryon number density of ordinary nuclear matter.
So, in addition to nucleons and electrons, neutron stars may be
expected to hide considerable populations of strangeness-carrying
$\Lambda$ hyperons, eventually accompanied with $\Sigma$ and $\Xi$
hyperon populations too \cite{glen85:b}. The total hyperon population
may be as large as 20\% \cite{glen85:b}.

\section{Kaons}

\begin{figure}[tb] 
\begin{center}
\leavevmode
\epsfig{figure=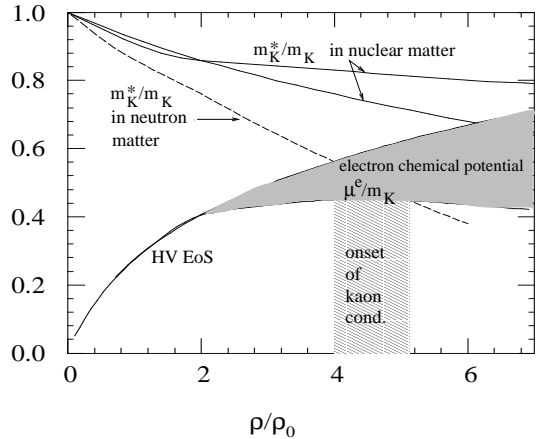,width=7.0cm,angle=0}
\caption[]{Effective kaon mass in nuclear \protect{\cite{mao99:a}} and
neutron star \protect{\cite{waas97:a}} matter (from
\cite{weber99:Ustron}).}
\label{fig:Kmass}
\end{center}
\end{figure}

Once the reaction $e^- \rightarrow K^- + \nu$ becomes possible in a
neutron star, it is energetically advantageous for the star to replace
the fermionic electrons with the bosonic $K^-$ mesons. The associated
chemical potentials must obey $\mu^e = \mu^{K^-}$.  Whether or not
this is actually fulfilled in neutron stars depends on the mass of the
$K^-$ in dense matter. Information about this is provided by the $K^-$
kinetic energy spectra extracted from Ni+Ni collisions at SIS
energies, as measured by the KaoS collaboration at GSI
\cite{barth97:a}.
 An analysis of the KaoS data shows that the attraction from
nuclear matter may bring the $K^-$ mass down to $m^*_{K^-}\simeq
200~\mev$ at $\rho\sim 3\, \rho_0$. For neutron-rich matter, the
relation
\begin{equation}
  m^*_{K^-}(\rho) \simeq m_{K^-} \left( 1 - 0.2 \, {{\rho} \over
  {\rho_0}} \right)
\label{eq:meff02}
\end{equation} was established \cite{li97:a,li97:b,brown96:a,brown97:a}, with
$m_K = 495$~MeV the $K^-$ vacuum mass.  Values around $m^*_{K^-}\simeq
200~\mev$ lie in the vicinity of the electron chemical potential,
$\mu^e$, in neutron star matter \cite{weber99:book,glen85:b} so that
the threshold condition for the onset of $K^-$ condensation,
$\mu^e(\rho) = m^*_{K^-}(\rho)$, could be fulfilled in the cores of
neutron stars.  The situation is graphically illustrated in figure
\ref{fig:Kmass}, with the uncertainties inherent in the calculations
indicted by the shaded regions.  Similar to the $\pi^-$ condensate, a
$K^-$ condensate would also soften the \eos and enhance the star's
neutrino luminosity \cite{tatsumi88:a,page90:a}.  The softening of the
\eos can be quite substantial, as pointed in \cite{li97:a,li97:b},
reducing the maximum neutron star mass from $\sim 2\, \msun$ to about
$1.5\, \msun$. Based on this, the existence of a large number of
low-mass black holes in the Galaxy was proposed \cite{brown94:a}.

\section{H-dibaryons}

Another boson that could make its appearance in the center of a
neutron star is the so-called H-dibaryon, a doubly strange six-quark
composite with spin and isospin zero, and baryon number two
\cite{jaffe77:a}. Since its first prediction in 1977, the H-dibaryon
has been the subject of many theoretical and experimental studies as a
possible candidate for a strongly bound exotic state.  In neutron
stars, which may contain a significant fraction of $\Lambda$ hyperons,
the $\Lambda$'s could combine to form H-dibaryons, which could give
way to the formation of H-matter at densities somewhere between $3\,
\rho_0$ \cite{glen98:a} and $6\, \rho_0$
\cite{tamagaki91:a,sakai97:a}, depending on the in-medium properties
of the H-dibaryon. H-matter could thus exist in the cores of
moderately dense neutron stars.  In \cite{glen98:a} it was pointed out
that H-dibaryons with a vacuum mass of about 2.2~GeV and a moderately
attractive potential in the medium of about $- 30$~MeV could go into a
Bose condensate in the cores of neutron stars if the limiting star
mass is about that of the Hulse--Taylor pulsar PSR~1913+16, $M=1.444\,
\msun$.  Conversely, if the medium potential were moderately
repulsive, say around $+ 30$~MeV, the formation of H-dibaryons may
only take place in neutron stars heavier than $M\sim 1.6\, \msun$. If
formed, however, H-matter may not remain dormant in neutron stars but,
because of its instability against compression could trigger the
conversion of neutron stars into hypothetical strange stars
\cite{sakai97:a,faessler97:a,faessler97:b}.

\section{Quark deconfinement}\label{ssec:deconf}

It has been suggested already back in the 1970's by a number of
researchers~\cite{fritzsch73:a,baym76:a,keister76:a,chap77:a,fech78:a,chap77:b}
that neutrons, protons plus the heavier constitutes ($\Sigma, \Lambda,
\Xi, \Delta$) can melt, creating the quark-gluon plasma state being
sought at the most powerful terrestrial heavy-ion colliders at CERN
and RHIC.  (Evidence for the creation of such matter in the framework
of CERN's Lead Beam Program has been claimed earlier this year
\cite{CERN00:a}.)  At present one does not know from experiment at
what density the expected phase transition to quark matter occurs, and
one has no conclusive guide yet from lattice QCD simulations.  From
simple geometrical considerations it follows that nuclei begin to
touch each other at densities of $\sim (4\pi r^3_N/3)^{-1} \simeq
0.24~\fmmt$, which, for a characteristic nucleon radius of $r_N\sim
1$~fm, is less than twice $\rho_0$.  Above this density, therefore,
one may expect that the nuclear boundaries of hadrons begin to
dissolve and the formerly confined quarks populate free states outside
of the hadrons.  Depending on rotational frequency and stellar mass,
densities as large as two to three times $\rho_0$ are easily surpassed
in the cores of neutron stars, so that the neutrons and protons in the
centers of neutron stars may have been broken up into their
constituent quarks by gravity \cite{weber99:topr}. More than that,
since the mass of the strange quark is only $m_s \sim 150$~MeV,
high-energetic up and down quarks will readily transform to strange
quarks at about the same density at which up and down quark
deconfinement sets in \cite{glen91:pt,kettner94:b}. It is therefore
three-flavor quark matter that could exist as a permanent component of
matter in the centers of neutron stars
\cite{glen97:book,weber99:book,weber99:topr,glen97:a}.  It is most
fascinating that observational astrophysicists may be able to spot
evidence for the existence of this novel phase of matter in the
neutron star data, as we shall see next.

\subsection{Anomalies in the spin-down behavior of isolated neutron stars}
\label{sec:ano1}

Whether or not quark deconfinement occurs in neutron stars makes only
very little difference to their static properties, such as the range
of possible masses and radii, which renders the detection of quark
matter in such objects extremely complicated. This turns out to be
strikingly different for rotating neutron stars (i.e.\ pulsars) which
develop quark matter cores in the course of spin-down.  The reason
being that as such stars spin down, because of the emission of
magnetic dipole radiation and a wind of electron-positron pairs, they
become more and more compressed.  For some rotating neutron stars the
\begin{figure}[tb] 
\parbox[t]{5.75cm} 
{\leavevmode
\epsfig{figure=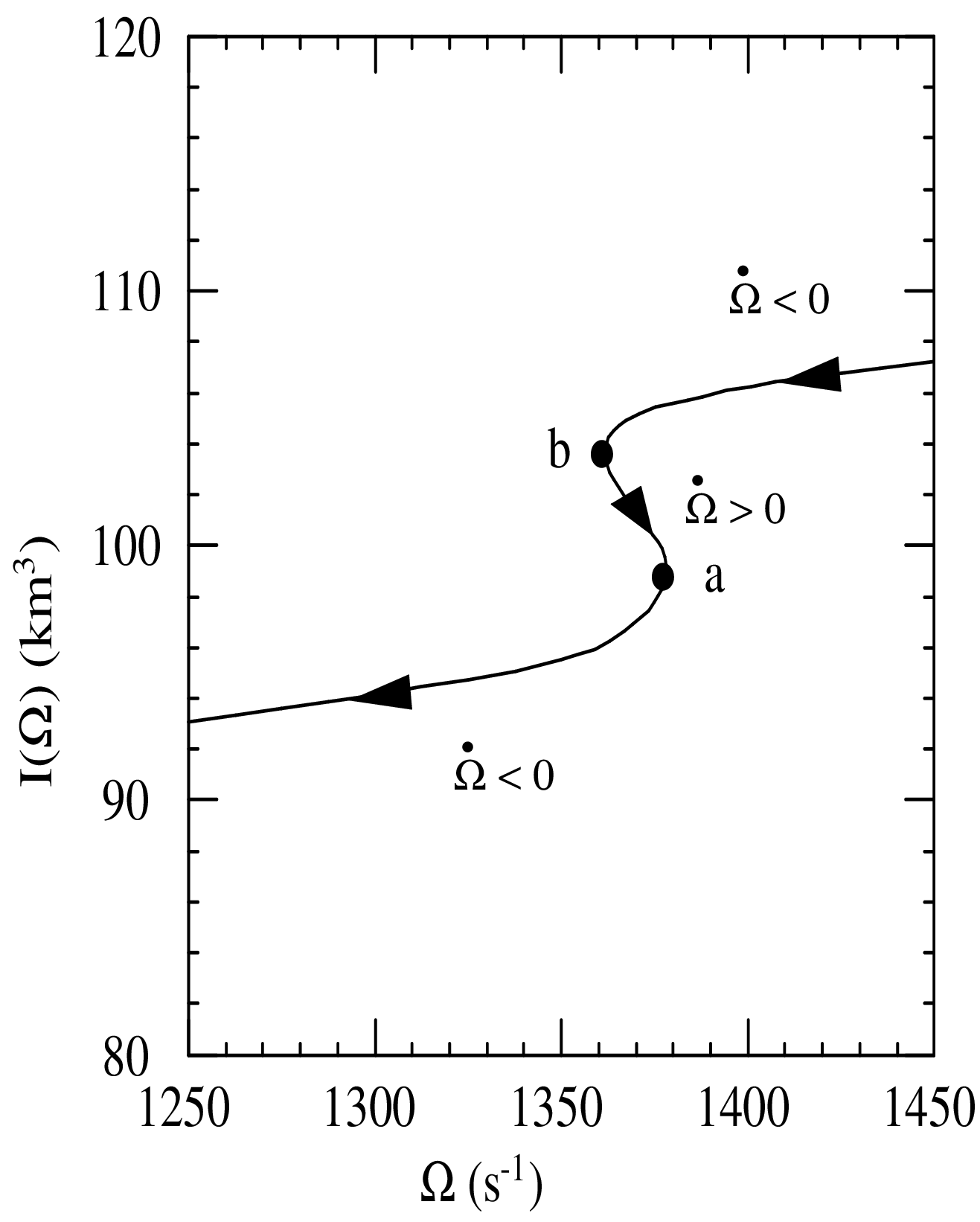,width=6.2cm,angle=-90}
{\caption[]{Moment of inertia versus frequency. The generation of
quark matter causes a ``backbending'' of $I$ for frequencies between $a$
and $b$ \protect{\cite{weber99:book,glen97:a}}.}
\label{fig:IOab}}}
\ \hskip0.50cm \ 
\parbox[t]{5.75cm} 
{\leavevmode
\epsfig{figure=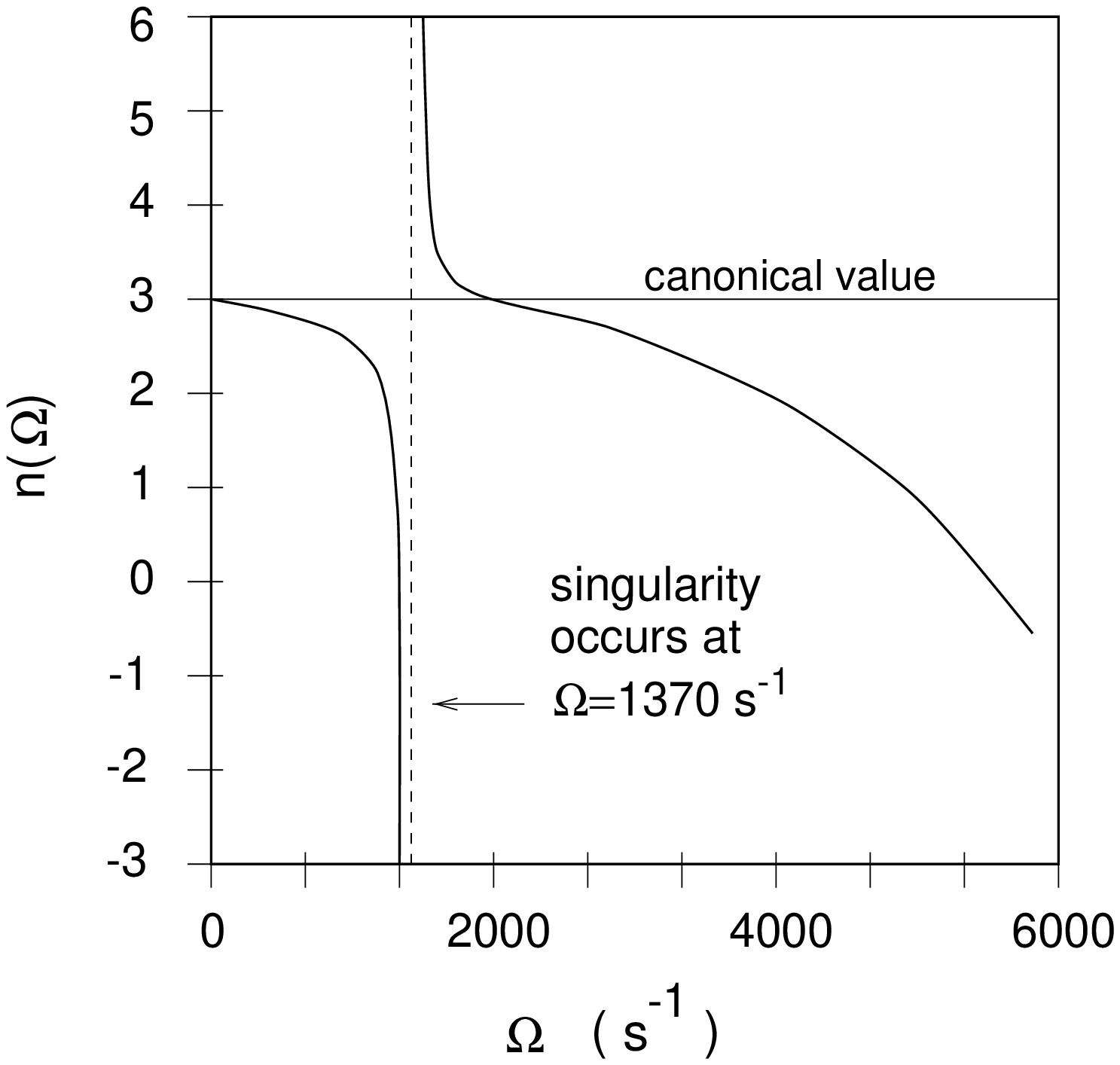,width=6.2cm,angle=-90}
\caption[]{Anomaly in braking index caused by the generation of quark matter.}
\label{fig:nvso}}
\end{figure}
mass and initial rotational frequency may be just such that the
central density rises from below to above the critical density for
dissolution of baryons into their quark constituents. This is
accompanied by a sudden shrinkage of the neutron star, which, as
calculated in \cite{glen97:a}, occurs at frequencies around 200~Hz
(angular velocities $\sim 1370~\secm$). This effects the star's moment
of inertia dramatically, as shown in figure~\ref{fig:IOab}. Depending
on the ratio at which quark and normal matter change with frequency,
the moment of inertia can decrease very anomalously, and could even
introduce an era of stellar spin-up lasting for $\sim 10^8$ years
\cite{glen97:a}.  Since the dipole age of millisecond pulsars is about
$10^9$~years, we may roughly estimate that about 10\% of the $\sim 25$
solitary millisecond pulsars presently known could be in the quark
transition epoch and thus could be signaling the ongoing process of
quark deconfinement!  Changes in the moment of inertia reflect
themselves in the braking index, $n$, of a rotating neutron star, as
can be seen from $(I'\equiv {\rm d}I/{\rm d}\Omega,~ I''\equiv {\rm
d}^2I/{\rm d}\Omega^2)$
\begin{equation}  
  n(\Omega) \equiv \frac{\Omega\, \ddot{\Omega} }{\dot{\Omega}^2} = 3
    - \frac{ 3 \, I^\prime \, \Omega + I^{\prime \prime} \, \Omega^2 }
    {2\, I + I^\prime \, \Omega} \, .
\label{eq:index}
\end{equation}  The right-hand-side of this expression reduces to the 
well-known canonical constant $n=3$ only if $I$ is independent of
frequency. Evidently, this is not the case for rapidly rotating
neutron stars, and it fails completely for stars that experience
pronounced internal changes which alter the moment of inertia
significantly. Figure \ref{fig:nvso} illustrates this for the neutron
star of figure~\ref{fig:IOab}.  Because of the changes in $I$, driven
by the transition into quark matter, the braking index deviates
dramatically from 3 at the transition frequency, when pure quark
matter is generated.  Such dramatic anomalies in $n(\Omega)$ are not
known for conventional neutron stars, because their moments of inertia
appear to vary smoothly with $\Omega$ \cite{weber99:book}. The future
astrophysical observation of such anomalies in the braking behavior
of pulsars may thus be interpreted as a signal for quark deconfinement
in neutron stars (``quark astronomy'').

\subsection{Anomaly in frequency distribution of neutron stars in low-mass
x-ray binaries}
\label{sec:ano2}

Accreting x-ray neutron stars provide a very interesting contrast to
the spin-down of isolated neutron stars. These x-ray neutron stars are
being spun up by the accretion of matter from a lower-mass ($M \ls 0.4
\msun$), less-dense companion.  If the critical deconfinement density
falls within that of the canonical pulsars, quark matter will already
exist in them but will be ``spun out'' of x-ray stars as their
frequency increases during accretion.  This scenario has been modeled
in \cite{weber00:a}, where it was found that quark matter remains
relatively dormant in the core of a neutron star until the star has
been spun up to frequencies at which the central density is about to
drop below the threshold density at which quark matter exists. As
known from above, this manifests itself in a significant increase of
the star's moment of inertia. The angular momentum added to a neutron
star during this phase of evolution is therefore consumed by the
star's expansion, inhibiting a further spin-up until the quark matter
has been converted into a mixed phase of matter made up of hadrons and
quarks.  Such accreters, therefore, tend to spend a greater length of
time in the critical frequencies than otherwise. There will be an
anomalous number of accreters that appear at or near the same
frequency. This is what was found recently with the Rossi x-ray Timing
Explorer (RXTE).  Quark deconfinement constitutes a most natural
explanation for this phenomenon, though alternative explanations are
possible too \cite{bildsten98:a,andersson00:a}.

\subsection{Diquark condensation and color superconductivity}

Recently it was discovered that instantons may cause strong
correlations between up and down quarks, which could give way to the
existence of colored diquark pairs in superdense matter
\cite{alford98:a,rapp98:a,rajagopal99:a}.\footnote{See also K.\
Rajagopals contribution elsewhere in this volume.}  These pairs would
form a Bose condensate in cold and dense quark matter.  Carrying color
charges, the condensate ought to exhibit color superconductivity
\cite{rapp98:a,rapp99:a}. Both the magnitude of the gap and the
critical temperature associated with the color superconductive phase
were estimated to be on the order of $\sim 100$~MeV
\cite{rapp98:a,rapp99:a}.  The implications of such tremendous gaps
for the magnetic fields of pulsars and their thermal evolution were
explored in \cite{blaschke99:a,alford99:a} and \cite{blaschke99:b},
respectively.  Color superconductivity may also be linked to the
sudden irregularities observed in the rotational frequencies of some
pulsars, known as glitches, as shown in the very recent study
\cite{alford00:a}.

\section{Absolutely stable quark matter}\label{ssec:ss}

So far we have assumed that quark matter forms a state of matter higher in
energy than atomic nuclei. This most plausible assumption, however, may be
quite deceiving \cite{bodmer71:a,witten84:a,terazawa89:a} because for a
collection of more than a few hundred $u,\, d,\, s$ quarks, the energy per
baryon ($E/A$) of quark matter can be just as well below the energy of
the most stable atomic nucleus, $^{56}\rm{Fe}$, whose energy per baryon number
is $M(^{56}{\rm Fe})c^2/56=930.4$~MeV, with $M(^{56}{\rm Fe})$ the mass of the
$^{56}$Fe atom.  A simple estimate indicates that for strange quark matter $E/A
= 4 B \pi^2/ \mu^3$, so that bag constants of $B=57.5~\mevt$ (i.e.
$\bag=145$~MeV) and $B=85.3~\mevt$ ($\bag=160$~MeV) place the energy per baryon
of such matter at $E/A=829$~MeV and 915~MeV, respectively
\cite{weber99:book,madsen88:a,madsen93:a,madsen97:bsky}.  Obviously, these
values correspond to quark matter which is absolutely bound with respect to
$^{56}$Fe. In this event the ground state of the strong interaction would be
\begin{figure}[tb]
\begin{center}
\leavevmode 
\epsfig{figure=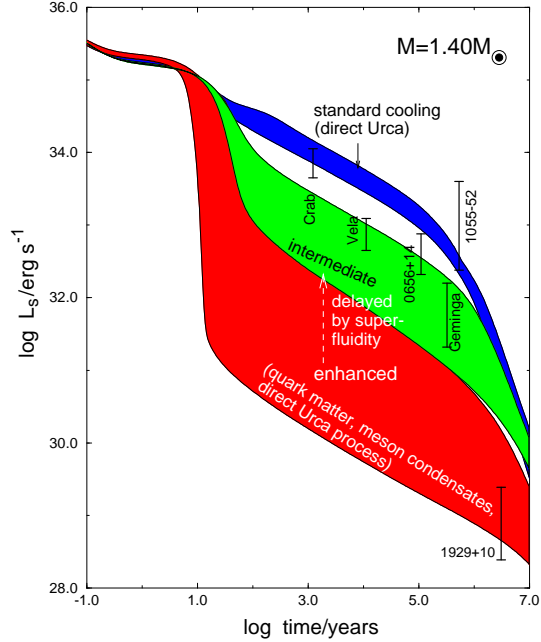,width=7.0cm}
\caption[]{Cooling behavior of a $1.4\,\msun$ neutron star based on
competing assumptions about the behavior of superdense matter. Three
distinct cooling scenarios, referred to as ``standard'',
``intermediate'', and ``enhanced'' (for details, see
\protect{\cite{weber99:book}}), can be distinguished. The band-like
structures reflects the uncertainties inherent in the underlying \eos
(cf.\ figure \ref{fig:eos}).}
\label{fig:cool} 
\end{center}
\end{figure}
strange quark matter (strange matter), made up of $u,\, d,\, s$
quarks, instead of nuclear matter.  This finding is one of the most
startling predictions of modern physics with far-reaching implications
for neutron stars, for all hadronic stellar configurations in figure
\ref{fig:cross} would then be only metastable with respect to stars
made up of absolutely stable 3-flavor strange quark matter
\cite{weber99:book,witten84:a,alcock86:a,haensel86:a,alcock88:a}.  If
this is indeed the case, and if it is possible for neutron matter to
tunnel to quark matter in at least some neutron stars, then in appears
likely that all neutron stars would in fact be strange stars
\cite{madsen88:a,madsen93:a,madsen97:bsky,friedman90:a,caldwell91:a}.
In sharp contrast to the other stars in figure \ref{fig:cross}, made
up of hadronic matter, possibly in phase equilibrium with quarks,
strange stars consist nearly entirely of pure 3-flavor quark matter,
eventually enveloped in a thin nuclear crust whose density is always
less than neutron drip density ($4\times 10^{11}~\gcmt$)
\cite{glen92:crust}.

The hypothetical, absolute stability of strange matter gives way to a
variety of novel objects which stretch from strangelets at the small
baryon number end, $A\sim 10^2$, to strange MACHOS and strange dwarfs,
to the compact strange stars at the high end, $A\sim 10^{57}$, where
strange matter becomes unstable against gravitational collapse
\cite{kettner94:b}.  The strange counterparts of ordinary atomic
nuclei are the strange nuggets, searched for in high-energy collisions
at Brookhaven (e.g.\ E858, E864, E878, E882-B, E886, E896-A), CERN
(Newmass experiment NA52), balloon-borne experiments (CRASH), and
terrestrial experiments (e.g.\ HADRON) \cite{weber99:book}.  (For a
recent review, see \cite{klingenberg99:topr} as well as this volume.)
One of the most distinguishing features of the compact strange stars
are their anomalously small radii which enables the whole branch of
compact strange stars to withstand very rapid rotation, down to
periods in the half-millisecond regime \cite{weber99:book}. In
contrast to this, only neutron stars very close to the limiting-mass
value may be able to withstand such rapid rotation.  An interesting
candidate for a small-radius ``neutron'' star could be
SAX~J1808.4-3658 whose radius, as argued in \cite{li99:a}, may be
significantly smaller than the canonical neutron star radius of $\sim
10$~km (see however \cite{glen00:a}).  Other distinguishing features
originate from the fact that the surfaces of bare strange stars are
made up of quark matter which allows for stellar luminosities as high
as several $10^{51}~\ergs$ \cite{usov98:a}.

\section{Cooling of neutron stars}

The predominant cooling mechanism of hot (temperatures of several
$\sim 10^{10}$~K) newly formed neutron stars immediately after
formation is neutrino emission, with an initial cooling time scale of
seconds. Already a few minutes after birth, the internal neutron star
temperature drops to $\sim 10^9$~K \cite{burrows86:a}. Photon emission
overtakes neutrinos only when the internal temperature has fallen to
$\sim 10^8$~K, with a corresponding surface temperature roughly two
orders of magnitude smaller. Neutrino cooling dominates for at least
the first $10^3$ years, and typically for much longer in standard
cooling (modified Urca) calculations.  Being sensitive to the adopted
nuclear \eosp, the neutron star mass, the assumed magnetic field
strength, the possible existence of superfluidity, meson condensates
and quark matter, theoretical cooling calculations, as summarized in
figure \ref{fig:cool}, provide most valuable information about the
interior matter and neutron star structure. The stellar cooling tracks
in figure \ref{fig:cool} are computed for a broad collection of \eoss
\cite{weber99:book}, including those shown in figure
\ref{fig:eos}. Knowing the thermal evolution of a neutron star also
yields information about such temperature-sensitive properties as
transport coefficients, transition to superfluid states, crust
solidification, and internal pulsar heating mechanisms such as
frictional dissipation at the crust-superfluid interfaces
\cite{schaab99:b}).

\section*{Acknowledgments}

This work was supported by the Deutsche Forschungsgemeinschaft (DFG).

\end{document}